\documentclass[twocolumn,showpacs,amsmath,amssymb]{revtex4}
\usepackage{graphicx}

\begin{document}

\title{Renormalization Group and Curved Spacetime}
\author{H. Matsueda}
\affiliation{Sendai National College of Technology, Sendai 989-3128, Japan}
\date{\today}
\begin{abstract}
We examine the role of curved geometry on renormalization group by means of image compression based on the singular value decomposition. By calculating course-grained images and their entanglement entropy, we find the anti-de Sitter space / conformal field theory correspondence hidden in the compression. The correspondence is originated from the conservation law for information. We discuss how one particular metric is automatically induced from various images. A formula for resolution of the course-grained images is also derived.
\end{abstract}
\maketitle

The renormalization group (RG) or the course graining is a fundamental concept in theoretical physics. Renormalization was originally considered to be a method for avoiding difficulties of divergences in quantum field theory, and RG was applied to critical phenomena in condensed matter physics. However, we may not still get to its heart, since the recent progress of RG requires more comprehensive understanding among entanglement, duality, holography, curved spacetime, and so on.

A key concept concerned with renormalization is the anti-de Sitter space / conformal field theory (AdS/CFT) correspondence~\cite{Maldacena,Gubser,Witten}. This is holographic correspondence between a quantum field theory in spatially $d$-dimension, CFT${}_{d+1}$, and general relativity on hyperbolic geometry in one-higher dimension, AdS${}_{d+2}$. Thus, the correspondence is closely related to quantum gravity, and at the same time it has been shown that the radial axis of AdS space corresponds to a renormalization flow parameter for CFT${}_{d+1}$ that is located at the boundary of noncompact AdS${}_{d+2}$ manifold~\cite{Fukuma}. The role of this RG flow on course-grained information is reflected in the entanglement entropy. The entropy of a subsystem in a CFT is proportional to the area of the minimal surface in AdS space surrounding the subsystem~\cite{Ryu}. This means that the total amount of the original information in the CFT is distributed to each RG layer.

In condensed matter physics, the real-space RG is a long-term topic. There are two main streams: one is Wilson's pioneering work for the Kondo problem~\cite{Wilson}, and the other is White's density matrix renormalization group method (DMRG)~\cite{White}. Recently, it has been found that the Wilson's RG controls the energy scale of low-lying excitations by exponentially bending the model space~\cite{Okunishi}. A systematic method for this energy-scale control is called hyperbolic deformation~\cite{Ueda}. On the other hand, DMRG is known to be a variational optimization method of the matrix product state (MPS) for gapped systems, and its extention to higher-dimensional systems is given by the tensor product state or the projected entangled pair state. The application of tensor networks to critical systems leads to the tree tensor network (TTN) and the multiscale entanglement renormalization anzats (MERA)~\cite{Vidal1,Vidal2}. They are spatially $(d+1)$-dimensional hierarchical networks that are compatible with real-space RG, and the network forms discrete AdS space. A crucial improvement in MERA is to introduce the disentangler transformation. In a viewpoint of the AdS/CFT correspondence, however, the network structure seems to play a role on succesive description of critical phenomena~\cite{Swingle}. Therefore, the hyperbolic geometry is a common ingredient in the two different RGs.

In order to explore the role of curved geometry on RG, we examine image compression based on the singular value decomposition (SVD) that is a core algorithm of DMRG and principle component analysis. Our approach is advantageous to extracting geometric information from images, since the color degrees of freedom can be regarded as the strength of the gravitational field. Furthermore, various field patterns are generated by only exchanging the images. We find the AdS/CFT correspondence hidden in the compression of any images. The entanglement entropy of course-grained images shows the universal scaling behavior in one-dimensional (1D) quantum systems at criticality. On the other hand, the course-grained image itself is devided into patches represented by linearly independent bases. Then, the RG flow of the patches forms discrete AdS space. We have a conjecture that the independency is related to the conformal symmetry. Based on the holographic entanglement entropy, we derive a formula for resolution of the course-grained images. The formula suggests the origin of the correspondence that comes from the conservation law for the amount of information.

Let us start with performing SVD to image data with the pixel size $M\times N$. We restrict our attention to grayscale images, but its extention to RGB data is straightforward. Then, the image data at pixel $(x,y)$, $\Psi(x,y)$, is represented by an integer ranging from $0$ to $255$. A white (black) pixel is assigned to be $255$ ($0$). We regard $\Psi(x,y)$ as a matrix, and $\Psi(x,y)$ is normalizaed to be $\Psi(x,y)\propto\psi(x,y)$ so that $\sum_{x,y}\psi^{2}(x,y)=1$. By applying the SVD to $\psi(x,y)$, we obtain
\begin{eqnarray}
\psi(x,y)=\sum_{l=1}^{L}U_{l}(x)\sqrt{\lambda_{l}}V_{l}(y), \label{svd}
\end{eqnarray}
with use of sigular values, $\lambda_{l}$, column unitary matrices, $U_{l}(x)$ and $V_{l}(y)$, and $L={\rm min}(M,N)$. We assume $\lambda_{1}>\lambda_{2}>\cdots>\lambda_{L}$ without loss of generality. 

The singular values are also the eigenvalues of the reduced density matrices defined by
\begin{eqnarray}
\rho_{X}(x,x^{\prime})&=&\sum_{y}\psi(x,y)\psi(x^{\prime},y)=\sum_{l}U_{l}(x)\lambda_{l}U_{l}(x^{\prime}), \\
\rho_{Y}(y,y^{\prime})&=&\sum_{y}\psi(x,y)\psi(x,y^{\prime})=\sum_{l}V_{l}(y)\lambda_{l}V_{l}(y^{\prime}).
\end{eqnarray}
In a viewpoint of DMRG, $X$- and $Y$-components of $\psi(x,y)$ behave as system and environment blocks of 1D systems. We thus expect that $\lambda_{l}$ decays exponentially as $l$ increases. Then, taking $m$-largest singular values is a good approximation for $\psi(x,y)$ ($m\ll L$)
\begin{eqnarray}
\psi(x,y)\sim\psi_{m}(x,y)=\sum_{l=1}^{m}U_{l}(x)\sqrt{\lambda_{l}}V_{l}(y).
\end{eqnarray}
A quality of the course-grained image, $\psi_{m}(x,y)$, is evaluated by the entanglement entropy defined by
\begin{eqnarray}
S_{m}=-\sum_{l=1}^{m}\lambda_{l}\log\lambda_{l}.
\end{eqnarray}
Note that $S_{m}$ picks up long range correlation more precisely than the truncation number in DMRG defined by $\epsilon_{m}=1-\sum_{l=1}^{m}\lambda_{l}$.

\begin{figure}[htbp]
\begin{center}
\includegraphics[width=10cm]{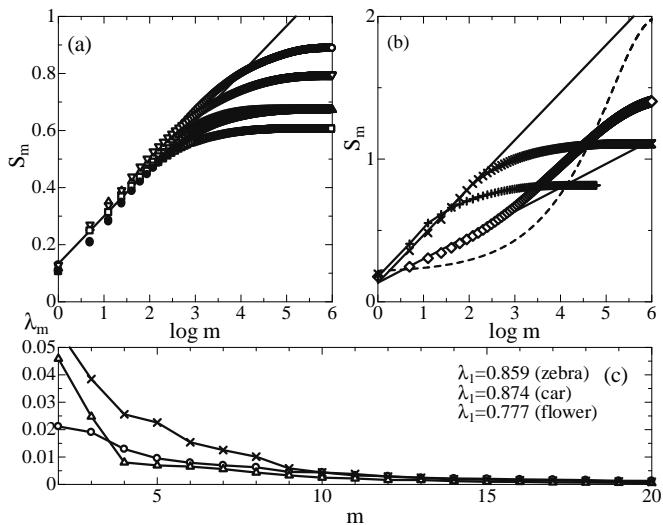}
\end{center}
\caption{$S_{m}$ for various images: (a) zebra ($\circ, 425\times 640$), car ($\bigtriangleup, 512\times 512$), my office ($\bigtriangledown, 512\times 384$), colleagues ($\bullet, 512\times 384$), cat ($\Box, 398\times 342$); (b) flower ($\times, 640\times 425$), two cats ($+, 398\times 342$), fractal tree ($\Diamond, 1024\times 768$), random dots (dashed line, $500\times 500$), (c) $\lambda_{m}$ for zebra, car, and flower.}
\label{scaling}
\end{figure}

\begin{figure}[htbp]
\begin{center}
\includegraphics[width=8.7cm]{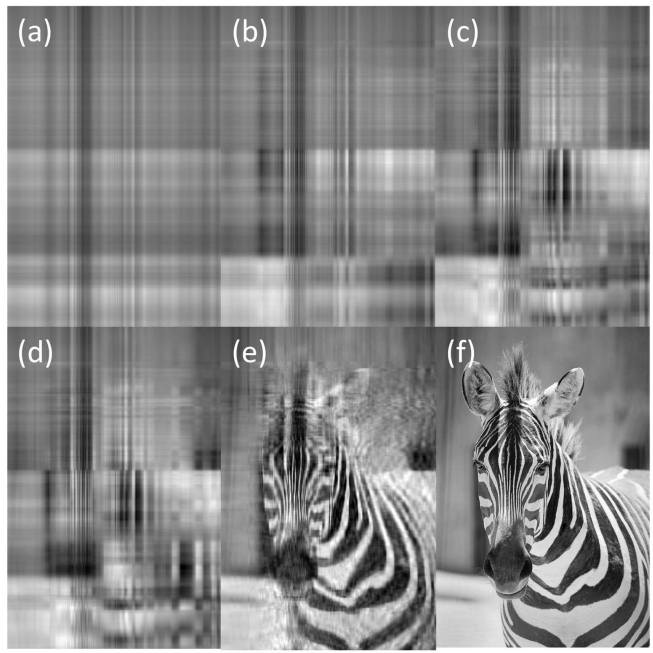}
\end{center}
\caption{Zebra image. (a) $m=1$, The black vertical line represents his nose and front leg. The lower half of this image shows oscillation of colors which comes from stripes of his body. The algorithm is going to replace the original image by one typical basis. (b) $m=2$, (c) $m=3$, (d) $m=4$, (e) $m=20$, (f) original image before compression ($m=425$).}
\label{zebra}
\end{figure}

Figure~\ref{scaling} shows $S_{m}$ (and $\lambda_{m}$) for various images as a function of $\log m$. Figures 2 and 3 present $\psi_{m}(x,y)$ for two different types of images (zebra and flower). Here, the pixel sizes of zebra and flower images are $425\times 640$ (horizontal $\times$ vertical) and $640\times 425$, respectively. Then we have $425$ independent eigenstates for both cases ($L=425$). In Fig.~\ref{scaling} (a), we clearly see that the various image data fit one straightline for small $m$ region:
\begin{eqnarray}
S_{m}=\frac{1}{6}\log m + \gamma,
\label{open}
\end{eqnarray}
with $\gamma=-\lambda_{1}\log\lambda_{1}\sim 0.13$. Here, we assume that the images are in the open boundary condition. This scaling behavior is also seen in the MPS formulation ($m$ represents the matrix dimension in this case) of the transverse-field Ising chain at criticality (central charge $c=1/2$), $H=-\sum_{i}(\sigma_{i}^{x}\sigma_{i+1}^{x}+\sigma_{i}^{z})$~\cite{Tagliacozzo}, and the $S=1$ $XXZ$ chain with the uniaxial anisotropy ($c=1$), $H=\sum_{i}(S_{i}^{x}S_{i+i}^{x}+S_{i}^{y}S_{i+1}^{y}+\Delta S_{i}^{z}S_{i+1}^{z})+D\sum_{i}(S_{i}^{z})^{2}$, with $\Delta=2.59$ and $D=2.3$~\cite{Huang}. They belong to the different universality classes, and thus the eq. (\ref{open}) is recognized to be universal for 1D quantum critical systems. When we take larger $m$ values, $S_{m}$ starts to deviate from the scaling line and approaches a saturated value. It depends on images how abrupt the saturation occurs. The saturated value tends to increase with $M$ and $N$, since the entropy measures the amount of information. Figure~\ref{zebra} (e) shows that the original image is almost recovered for $m\sim 20$ ($\log m\sim 3$) where the deviation starts.

\begin{figure}[htbp]
\begin{center}
\includegraphics[width=9cm]{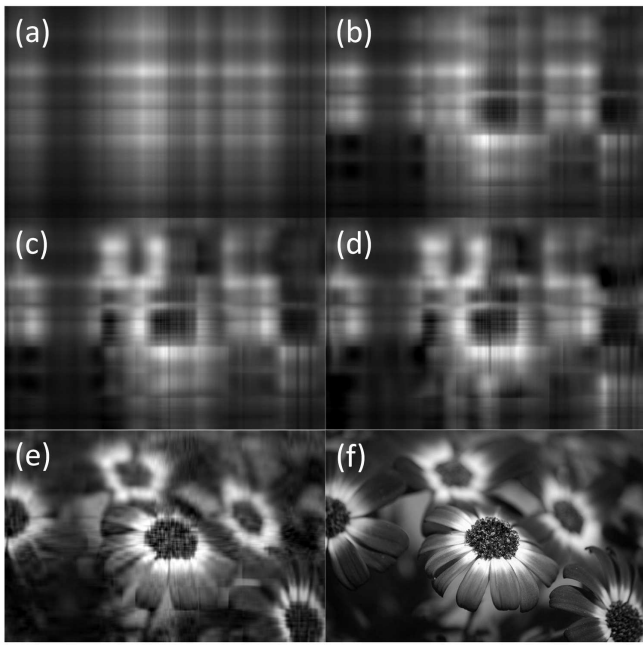}
\end{center}
\caption{Flower image. (a) $m=1$, (b) $m=2$, (c) $m=3$, (d) $m=4$, (e) $m=20$, (f) original image before compression ($m=425$).}
\label{flower}
\end{figure}

Figure~\ref{scaling} (b) shows exceptional cases that do not obey the eq. (\ref{open}). For the flower image (cross symbols), the entropy for small $m$ region has a gradient two times larger than the previous examples:
\begin{eqnarray}
S_{m}=\frac{1}{3}\log m + \gamma,
\label{periodic}
\end{eqnarray}
where the residual entropy $S_{1}=\gamma$ is still unchanged. According to the CFT${}_{1+1}$, the difference of the gradient in the entropy comes from boundary conditions. In the present case, the scale invariance among flowers seems to induce the change in the gradient (see Fig.~\ref{flower}). For the tree image in Fig.~\ref{scaling} (b), the gradient of the entropy jumps from $1/6$ to $1/3$ at around $m\sim 20$. This comes from the fractal nature that the branches and the trunk have quite similar structures. The branch structure can not be seen for $m<20$, while their fine structures appear almost abruptly as we increase $m$. According to the Zamolodchikov's $c$-theorem, the fractal case is special, since the $c$-function corresponding to the gradient of $S_{m}$ does not change monotonically~\cite{Zamolodchikov}.

\begin{figure}[t]
\begin{center}
\includegraphics[width=9cm]{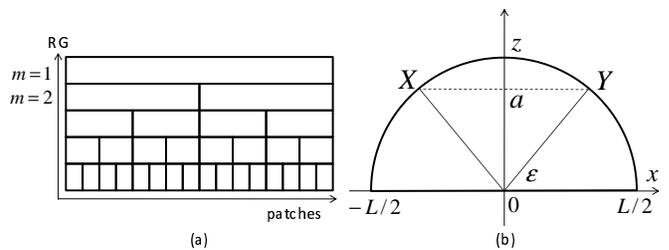}
\end{center}
\caption{(a) Discrete AdS space, (b) Geodesic curve on continous AdS space (CFT is defined on the $x$-axis).}
\label{AdS}
\end{figure}

Let us next look at the course-grained images that produce the eq. (\ref{open}). We find that the image space is devided into smaller patches as $m$ increases. For instance in Fig.~\ref{flower} (a) with $m=1$ the image is characterized by one horizontal line, while in Fig.~\ref{flower} (b) with $m=2$ we clearly see discontinuity between upper and lower halves marked by a left arrow. The discontinuity comes from a fact that the upper and lower halves are represented by linearly independent basis states. This point may be formulated more precisely by the multi-resolution analysis, and then we conjecture that the scale factor of the mother wavelet plays a role on the conformal symmetry. In the present case, the poistion of the partitioning is automatically selected so that the linear independency is optimized by DMRG. If we regard these patches as independent lattice sites, the lattice plus the RG axis forms the discrete AdS space shown schematically in Fig.~\ref{AdS} (a). The construction of our AdS space is analogous to that of the MERA network with disentangler. TTN is also similar to the AdS space, but the tensors are still entangled. Therefore, the disentangling processes are crucial for generating the AdS space.

Note that Fig.~\ref{AdS} (a) is not a oversimplifyed view to analyze the scaling relation even quantitatively. The continous AdS metric has various expressions. For instance,
\begin{eqnarray}
ds^{2}=\frac{l^{2}}{z^{2}}\left(dz^{2}+dx^{2}\right)=\big( d\tau \log\alpha \big)^{2} + \big( \alpha^{-\tau/l}dx \big)^{2}
\end{eqnarray}
with curvature $l$ and $z=\alpha^{\tau/l}l$. The factor $\alpha$ determines how many patches merge together, but this transformation guarantees that the most simple expression, $ds^{2}=(l^{2}/z^{2})(dz^{2}+dx^{2})$, is essential.

Let us derive the resolution of the course-grained images by calculating the holographic entanglement entropy within a continuous approximation (see Fig.~\ref{AdS}(b))~\cite{Ryu}: 
\begin{eqnarray}
S=\frac{\gamma}{4G},
\end{eqnarray}
where $\gamma$ represents the geodesic distance between $X$ and $Y$ on the same RG layer with $m$, and $G$ is the Newton constant. The geodesic equation is given by $(d^{2}x^{\lambda}/dt^{2})+\Gamma_{\mu\nu}^{\lambda}(dx^{\mu}/dt)(dx^{\nu}/dt)=0$ with $x^{\mu}=(x^{1},x^{2})=(x,z)$ and the Christoffel symbol $\Gamma_{\mu\nu}^{\lambda}$. A solution is a half cycle: $(x,z)=(L/2)(\cos\theta, \sin\theta)$, where the angle $\theta$ obeys the following relation $d\theta/dt=\sin\theta$ ($\epsilon\le\theta\le\pi-\epsilon$). We introduce a parameter $a=(L/2)\sin\epsilon$ instead of $\epsilon$. The geodesic distance $\gamma$ is then calculated by
\begin{eqnarray}
\gamma =\!\!\int_{\epsilon}^{\pi-\epsilon}\!\!\frac{l}{z}d\theta\sqrt{\left(\frac{\partial z}{\partial\theta}\right)^{\!\! 2}+\left(\frac{\partial x}{\partial\theta}\right)^{\!\! 2}} = l\log\left(\frac{1+\cos\epsilon}{1-\cos\epsilon}\right),
\end{eqnarray}
and the entropy is given by
\begin{eqnarray}
S=\frac{1}{6}c\log \left[\frac{\left(L+\sqrt{L^{2}-(2a)^{2}}\right)^{2}}{(2a)^{2}}\right],
\end{eqnarray}
where $c=3l/2G$~\cite{Henningson}. Usually, we take the limit $a\rightarrow 0+$ in order to calculate the entropy of the original quantum systems. Then, we obtain $S=(c/3)\log(L/a)$~\cite{Calabrese}. On the other hand, here we are going to approach $2a\rightarrow L$, and the entropy should be comparable with the scaling relation (\ref{open}). Thus, we obtain
\begin{eqnarray}
m=\left(\frac{L+\sqrt{L^{2}-(2a)^{2}}}{2a}\right)^{\!\! 2c}.
\end{eqnarray}
Finally, the resolution of a course-grained image for a fixed $m$ value, $R_{m}$, is given by
\begin{eqnarray}
R_{m}=\frac{|X-Y|}{L}=\frac{1}{L}2\sqrt{\left(\frac{L}{2}\right)^{\!\! 2}-a^{2}}=\frac{m^{\frac{1}{c}}-1}{m^{\frac{1}{c}}+1}.
\end{eqnarray}
This formula is consistent with numerical data, when we take $c=1$. Here $R_{2}=0.333$, $R_{10}=0.818$, $R_{20}=0.905$, and $R_{40}=0.951$. According to this trend, we have already mentioned that the data with $m=20$ almost reproduce the original one except for very fine details. We have confirmed that the compressed data with $m=10$ starts to reproduce the original one, but fine structures are still lost. The reasonable correpondence between the holographic theory and the numerical data also supports the existence of the AdS/CFT duality.

\begin{figure}[t]
\begin{center}
\includegraphics[width=9.5cm]{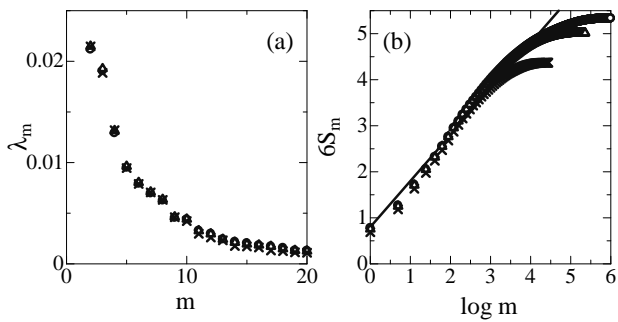}
\end{center}
\caption{(a) $\lambda_{m}$ and (b) $S_{m}$ for the zebra images with different pixel sizes: $425\times 640$ pixels (open circles), $212\times 320$ pixels (open triangles), and $85\times 128$ pixels (crosses).}
\label{resolution}
\end{figure}

Based on the duality of coupling parameters between AdS and CFT, we consider the reason why the Gaussian universality class, $c=1$, appears. The zebra image represents an 'ordered' pattern induced by strong interactions in this image space. Then, we can interplete the present results as a consequence of free bosons in the CFT side. We can exclude the possibility of taking another $c$ values such as the Ising universality class $c=1/2$ and other unitary discrete series, since the images do not have any $Z_{2}$ symmetry. Furthermore, we always obtain the positive residual entropy $\gamma$, and this is due to weak topological nature of 1D systems~\cite{Kitaev,Levin}. When we look at the zebra image, he has various stripe patterns: The stripes of his body are wide, his face has somehow narrower stripes than his body, and his hair is more complicated. In particular, his hair represents high frequencies, which correspond to low-energy behavior in the CFT. Actually, when we increase resolution of the zebra image by taking larger $M$ and $N$ values, fitting to the scaling relation becomes much better for larger $m$ values (see Fig.~\ref{resolution}). This data clearly show the infrared cut-off in the CFT.

Summarizing, the core of the AdS/CFT correspondence lies in the conservation law for information between the quantum and classical sides: The high probability bases of the reduced density matrix in the quantum side represent course-grained geometry in the classical side (see also Fig.~\ref{scaling} (c)). The curvature of the classical space determines the amount of information that should remain after RG transformation.

The author would like to thank N. Shibata, K. Totsuka, K. Okunishi, T. Nishino, I. Maruyama, H. Ueda, and M. Wada for discussions and comments.

\end{document}